\documentclass[useAMS,usegraphicx]{mn2e}
\usepackage{amsmath}

\newcommand{\ApJ}[3]    {\mbox{#3, ApJ,~#1,~#2}}

\newcommand{\MNRAS}[3]  {\mbox{#3, MNRAS,~#1,~#2}}
\newcommand{\Nature}[3] {\mbox{#3, Nature,~#1,~#2}}
\newcommand{\NewA}[3]   {\mbox{#3, NewA,~#1,~#2}}

\newcommand{\PhRevB}[3] {\mbox{#3, Phys.~Rev.~B#1,#2}}
\newcommand{\PhRevD}[3] {\mbox{#3, Phys.~Rev.~D#1,#2}}
\newcommand{\PhRep}[3]  {\mbox{#3, Phys.~Rep.,~#1,~#2}}

\def\ap3m{AP$^3$M}

\def\h0{$h_0$}
\def\H0{$H_0$}

\def\Msun{${\rm M}_{\odot}$}

\def\d{{\rm ~d}}

\def\ln{{~\rm ln}}

\def\ea{et~al.~} 
\def\Exp{{~\rm exp}}

\def\lesssim{\mathrel{\hbox{\rlap{\hbox{\lower4pt\hbox{$\sim$}}}\hbox{$<$}}}}
\def\grtsim{\mathrel{\hbox{\rlap{\hbox{\lower4pt\hbox{$\sim$}}}\hbox{$>$}}}}

\newcommand{\bh}    {\bullet}

\begin{document}
\title[MBH remnants of the first stars III: observational signatures
  from redshift evolution]{Massive black hole remnants of the
  first stars III: observational signatures from the past}

\author[Islam R.R., Taylor J.E. \& Silk J.]
       {Ranty R. Islam\thanks{Email: rri@astro.ox.ac.uk}, James E. Taylor\thanks{PPARC fellow} and Joseph Silk\\
       {Astrophysics, Denys Wilkinson Building, Keble Road, Oxford, OX1 3RH, UK}}

\date{Received ...; accepted ...}
\date{Accepted 2004 July 13. Received 2004 July 13; in original form 2003 October 24}

\maketitle

\begin{abstract}
The first stars forming in minihaloes at redshifts greater than 20 may
have been very massive and could have left behind massive black hole (MBH)
remnants. In previous papers we investigated the hierarchical merging
of these `seed' MBHs and their associated haloes, using a semi-analytical
approach consisting of a hierarchical merger tree algorithm and
explicit prescriptions for the dynamics of merged substructure inside a
larger host halo following a merger. We also estimated accretion
luminosities for these MBHs and found them to be consistent with
observations of ultra-luminous X-ray point sources.
Here we compute the strength of gravitational wave events as MBHs
merge to form the more massive black holes that we predict reside
in galaxy haloes today. If MBHs merge efficiently, we predict
that as many as $10^4$--$10^5$ events per year may fall within
the sensitivity limits of the proposed LISA gravitational wave observatory.
The collapse of the first massive stars to form MBHs may also be
accompanied by gamma-ray bursts (GRBs). If this is the case and if
GRBs are observable out to the redshifts of first star formation,
we predict that about $10^5$--$10^6$ GRBs per year could be detected.
As merging MBH binaries reach their last stable orbits before final
coalescence a fraction of the gravitational wave
energy may be released as a pulse of gamma rays
(for instance, through interaction
with material enveloping a merging MBH binary). 
This fraction has to be larger than about $10^{-2}$ for MBH mergers 
to account for some beamed GRBs, and greater than $10^{-6}$ for the gamma 
rays to be detectable out to cosmological distances with upcoming GRB detector missions.
\end{abstract}

\section{Introduction}
There is increasing speculation that the first generation of stars
in the universe may have been extremely massive, and that some of
these objects could have collapsed directly to massive black holes (MBHs)
at the end of a brief stellar lifetime. If these objects do form
in the early universe, then they may provide a unique opportunity
to study primordial star formation, through their gravitational
wave or gamma-ray emission.

Recent semi-analytic \cite{hutchings02,fuller00,tegmark97}
and numerical studies \cite{bromm02,abel00} suggest that the
first stars in the universe formed inside dense baryonic cores,
as they cooled and collapsed within dark matter haloes at very high redshift.
For a standard $\Lambda$CDM cosmology, these {\em minihaloes} are
estimated to have had masses in the range
$M_{min} \sim 10^5$--$10^6 ~h^{-1}$ \Msun,
and to have collapsed at redshifts $z_{\rm collapse} \sim 20$--$30$
or higher.
Since these first star-forming clouds contained essentially no metals,
gas cooling would proceed much more slowly in these systems
than in present-day molecular clouds. As a result, they may have
collapsed smoothly and without fragmentation, producing dense cores much
more massive than the proto-stellar cores observed in star-formation
regions today. Assuming nuclei within these cores accreted the surrounding
material efficiently, the result would be a first generation of protostars
with masses as great as $10^3$ \Msun\ \cite{bromm02,omukai01}.

As yet, nothing definite is known about the subsequent evolution
of such objects. However, their large masses would probably result
in many of them ending up as MBHs with little intervening
mass loss -- for systems in this mass range, gravity is so strong
that there is no ejection of material from the system in a final
supernova bounce \cite{heger02}.
This high-redshift population of MBHs is interesting as a source
of seeds for the formation of the supermassive black holes (SMBHs) 
seen at the centres
of galaxies at the present-day. Furthermore, the initial
collapse of these objects or mergers between them might
be detectable through the resulting gamma-ray or gravitational wave
emission. This would provide a direct test of the physics
of primordial star formation and the high-redshift universe.

In a previous paper \cite{islam_I} we used a semi-analytical model
of galaxy formation to follow the evolution of MBHs as they
merge together hierarchically, along with their associated haloes.
The code we used combines a Monte-Carlo algorithm to generate halo
merger trees with analytical descriptions for the main dynamical processes --
dynamical friction, tidal stripping, and tidal heating --
that determine the evolution of merged remnants within a galaxy halo.
We introduce seeds into the code by assuming that in each minihalo
forming before a redshift $z_{\rm collapse}$, a single  MBH forms as the
end result of primordial star formation.

For our computations, we considered four different sets of values for
the parameters $\nu_{pk}$ and $M_{\bh,seed}$, which fix the abundance
and the mass of the seeds respectively. These values are summarised in
table \ref{tab:inits}.
Our choice of a seed MBH mass of 260 \Msun\ is motivated by the result
of Heger (2002) \nocite{heger02} that massive stars above this mass
will not experience a supernova at the end of their lives, but instead
will collapse directly to a MBH of essentially the same mass.
\begin{table}
  \begin{center}
    \caption{The mass of seed MBHs, and the height of the peaks in the initial
    density field in which they formed, for the four models considered.
    The collapse redshift $z_{\rm collapse}$ is
    the epoch when peaks of height $\nu_{pk}$ first
    cross the cooling threshold (see paper I).}
    \begin{tabular}{cccc} \\    \hline \hline \label{tab:inits}
      Model & $M_{\bh,seed}$ & peak height $\nu_{pk}$ & $z_{\rm collapse}$\\ \hline
      A & 260 \Msun\ & 3.0 & 24.6\\
      B & 1300 \Msun\ & 3.0&  24.6\\
      C & 260 \Msun\ & 2.5 & 19.8\\
      D & 260 \Msun\ & 3.5 & 29.4\\ \hline \hline
    \end{tabular}
  \end{center}
\end{table}

In paper I we investigated the abundance of MBHs in present-day galaxies
as a result of this process of hierarchical merging and dynamical
evolution. For these MBHs we estimated the X-ray accretion luminosities
using two
different accretion models and found that they could well account
for many of the ultra-luminous off-centre X-ray point sources that
are observed in local galaxies\cite{islam_II}.

In this paper we investigate how our model could be tested at high
redshifts. In particular, we predict the number of gravitational
wave events arising from MBH mergers that should be detectable
with the upcoming LISA mission. Conversely, observations of
gravitational wave events could then be used to constrain the
merging history of MBHs and provide limits on the abundance of
these objects at cosmological distances and to very high redshifts
(see also Haehnelt (1994) \nocite{haehnelt94}, Menou, Haiman \&
Narayanan (2001) \nocite{menou01}). If the collapse of the first
massive stars to form MBHs is accompanied by gamma-ray bursts
(GRBs), then observations of GRBs may allow us to directly probe
the epoch of first star formation at redshifts larger than 20.
Here we estimate the number of GRB events that could be observed
per year if GRBs are detectable out to the redshifts of first star
formation.

The structure of the paper is as follows.
In section \ref{sec:G_waves} we compute the number of MBH mergers and
the number of associated gravitational wave events.
Section \ref{sec:GRBs} considers
the expected number of GRBs from collapsing massive population III
stars, as well as the possibility of gamma-ray emission in the wake of
MBH mergers.
We discuss and summarise our findings in section \ref{sec:summary}
\section{Gravitational waves from SMBH-MBH mergers}
\label{sec:G_waves}

There are essentially two types of gravitational waves
that are emitted at different stages in the evolution of the scenario
we are considering.
The first of these are `burst' signals associated with the collapse of
massive population III stars into MBHs. In our model, these are events
occurring at very high redshifts $>$ 20 for a cosmological abundance of
massive population III stars, so we expect a stochastic
gravitational wave background to be generated
\cite{schneider00,araujo02}.
Although statistical limits can be imposed, the particular number and
strength of burst signals created in this way does depend on the
initial number and masses of collapsing stars and the collapse
mechanism.

In contrast periodic signals, albeit with changing frequency, are
emitted during the subsequent merger of MBHs to form more massive
BHs. Compared to the burst signals from collapsing population III
stars, the mergers of MBHs produce stronger waves due the higher
total mass of the gravitationally interacting system. As a result
gravity waves could be detected for individual merger events. The
signals from merger events can be further subdivided into signals
associated with the inspiral phase, actual coalescence and
subsequent ringdown \cite{flanagan98}. Of these, the inspiral
phase is the longest and thus most likely to detect. In the
following we will be exclusively concerned with the gravitational
waves  emitted during this phase. These are also referred to as
`chirp' signals, due to the fact that during inspiral the distance
between the BHs in the merging binary decreases continuously as
energy is lost through gravity waves, resulting in the frequency
of emitted waves rapidly sweeping upwards.


\subsection{Gravitational Wave Amplitude}
For any measurement of gravitational waves the most important pieces
of information
are the frequency and some measure of their amplitude, both of which
will be functions of time.
Because of the spin-2 nature of gravity, the first non-zero and largest
term of the tensor
describing the gravity wave induced perturbation of space-time is the
quadrupole moment.
This can be described by an overall dimensionless strain amplitude $h_s$.
For periodic signals such as those emitted by a binary black hole
system, this is given by \cite{thorne89}
\begin{equation}
    h_s = 8 \left(\frac{2}{15}\right)^{1/2} \frac{G^{5/3}\mu_{12}}{c^4 D(z)}
(\pi M_{12}f_{gw})^{2/3}
\end{equation}
where $f_{gw}$ is the frequency of gravitational waves, which in this case is
just twice the orbital frequency of the binary; $M_{12}= M_1 + M_2$ and $\mu_{12} = M_1 M_2/M$, G is Newton's constant
and c the speed of light. $D(z)$, the distance to the binary,
can be obtained by integrating the differential distance redshift relation
\begin{equation}
  D(z) = \frac{c}{H_0} \int_0^z \frac{{\rm d}z'}{(\Omega_m (1+z')^3 +
    \Omega_{\Lambda})^{-1/2}}
\end{equation}
The effective strain amplitude measured at some characteristic
frequency $f_c$ is further enhanced proportionally to the square root
of the number of periods, n, that the binary emits at or near that frequency.
For an inspiralling binary system whose frequency sweeps up, or
`chirps',in frequency (cf.~below), $n \propto \mu_{12}^{-1/2} M_{12}^{-1/3}
f_c^{-5/6}$ and the characteristic strain amplitude, $h_c$ is thus \cite{thorne89,nakamura97}
\begin{equation} \label{eq:hc}
h_c = 2.94 \times 10^{-19} \left(\frac{M_{chirp}}{10^3 {\rm M}_{\odot}}\right)^{5/6}
\left(\frac{f_c}{{\rm Hz}}\right)^{-1/6} \left(\frac{D(z)}{10 {\rm
Mpc}}\right)^{-1}
\end{equation}
where $f_c$ is the frequency of the wave signal, $r$ the distance from
the merging binary and the `chirp' mass $M_{chirp}$ accounts for the masses $M_1$
and $M_2$ of the binary constituents
\begin{equation}
M_{chirp} = \frac{(M_{1} M_{2})^{3/5}}{(M_{1} + M_{2})^{1/5}} =
\mu_{12}^{3/5} M_{12}^{2/5}
\end{equation}

Particularly at late times, most mergers in our model
are between a massive central MBH and much lighter seed MBHs, so we can
consider the problem in the limit that one of the merging binary
constituents is much larger than the other.
If we assign $M_1$ and $M_2$ to the masses of the central and
inspiralling MBH respectively, $M_1 \gg M_2$ and
\begin{equation}
    M_{chirp} \approx M_1 x^{3/5}
\end{equation}
where $x = M_2/M_1$ denotes the mass ratio between the two masses.

The largest frequency at which a gravitational wave source can emit is
determined by its light radius which in turn is dependent on its mass,
and so in this case $f_{max} \approx 10 {\rm ~Hz} (10^3
{\rm M}_{\odot}/M_1)$. A lower limit on the characteristic wave amplitude is
then

\begin{equation}\label{eq:hcmin}
    h_c \approx 2 \times 10^{-19} \sqrt{x} \left(\frac{M_1}{10^3
{\rm M}_{\odot}} \right) \left( \frac{D(z)}{10 {\rm Mpc}}\right)^{-1}
\end{equation}


\subsection{Frequency of gravitational Waves}
The proposed LISA (Laser Interferometer Space Antenna) gravitational
wave observatory will be the only instrument
capable of detecting gravity waves in the typical frequency range
generated in the inspiral phase of a merging binary of massive black
holes.
A rough estimate of the maximum frequency of waves emitted by the
merging binary is
\begin{equation}\label{eq:GWfmax}
f_{max} \sim 10 {\rm ~Hz} \frac{10^3 {\rm M}_{\odot}}{M_{1}}
\end{equation}
where $M_1$ is the more massive of the binary constituents.

For a binary on circular orbit, the frequency of the chirp signal can
be described analytically. If the orbital radius is
$a_0$ at time $t_0 = 0$, then the time for the binary constituents to
spiral into each other is \cite{misner73}
\begin{equation}\label{eq:GWinspiraltime}
\tau_0 = 0.665 \left(\frac{M_{12}}{10^3 {\rm M}_{\odot}}\right)^{-2}
\left(\frac{\mu_{12}}{10^3 {\rm M}_{\odot}}\right)^{-1} \left(\frac{a_0}{10^8 {\rm
m}}\right)^4 {\rm yr}
\end{equation}
The frequency of the waves emitted at a later time $t$ is
\begin{equation}
    f_{gw} = 3.11 \times 10^{-3} \left(\frac{M_{chirp}}{10^3
{\rm M}_{\odot}}\right)^{-5/8} \left(\frac{\tau_0 - t}{{\rm
yr}}\right)^{-3/8} {\rm Hz}
\end{equation}
Including the redshift $z$, the waves are then detected by LISA with a
frequency
\begin{equation}
    f_{gw} = \frac{3.11 \times 10^{-3}}{(1+z)} \left(\frac{M_{chirp}}{10^3
{\rm M}_{\odot}}\right)^{-5/8} \left(\frac{\tau_0 - \frac{t}{1 + z}}{{\rm
yr}}\right)^{-3/8} {\rm Hz}
\end{equation}

For the case that $M_1 \gg M_2$, $\tau_0$ and
$f_{gw}$ are approximately given by

\begin{equation}
\tau_0 \approx \frac{0.665}{x} \left(\frac{M_1}{10^3 {\rm M}_{\odot}}\right)^{-3}
\left(\frac{a_0}{10^8 {\rm m}}\right)^4 {\rm yr}
\end{equation}

\begin{equation}
    f_{gw} \approx \frac{3.11 \times 10^{-3}}{(1+z)} \left(\frac{M_1}{10^3
{\rm M}_{\odot}}\right)^{-5/8} \left(x\frac{\tau_0 - \frac{t}{1 + z}}{{\rm
yr}}\right)^{-3/8} {\rm Hz}
\end{equation}

\subsection{MBH Merger Efficiency} \label{sec:merger_efficiency}
We have outlined above the main characteristics of
the gravitational wave signals
received from individual merger events. The formulae given estimate
the amplitude and frequency {\it averaged} over all inclinations of
the merger orbital plane.

For a merger event to actually occur the MBHs have to come within
a distance where gravitational waves can efficiently reduce the
orbital energy and allow the MBHs to spiral together and coalesce.
However, this only happens at MBH separations that are very much
smaller than the ones we are considering here (of order 10 - 100
pc).

For a MBH binary on a circular orbit to coalesce through gravitational
wave emission within a Hubble time,
the MBH separation, $a$, needs to satisfy \cite{peters64}
\begin{eqnarray}
  a ~\leq ~a_{gw} &=&  \left(\frac{256 ~t_{hubble} ~G^3 ~\mu_{12}
    ~M_1^2}{5~c^5}\right)^{1/4}\\ \nonumber
&\approx&  4.5\times10^{-6} \left(\frac{h}{0.7}\right)^{-1/4}
  \left(\frac{M_1}{260 ~M_{\odot}}\right)^{3/4} \\ \nonumber
  & & \times \left(\frac{M_2(M_1 +
  M_2)^2}{M_1^2}\right)^{1/4}~{\rm pc}
\end{eqnarray}

At distances larger than this the common environment of the MBHs
determines whether and how quickly the MBHs will be driven towards
each other.
If two MBHs come within a distance
\begin{eqnarray}
  a &\lesssim& a_h = \frac{G ~M_2}{4~\sigma_c^2} \\ \nonumber
  &\approx& 7.3\times10^{-2}
  \left(\frac{M_2}{260 M_{\odot}}\right)\left(\frac{\sigma_c}{200
  ~{\rm km~s^{-1}}}\right)^{-2} ~{\rm pc}
\end{eqnarray}
they form a `hard' binary. Interactions with stars in the binary's
vicinity carry away energy from the binary with the result that the
binary separation shrinks -- the binary is said to `harden'. The time
scale on which this happens is given by
\begin{equation}
 t_h = \frac{\sigma_c}{G ~\rho_{\star} ~a ~H_h}
\end{equation}
where $H_h$ is the dimensionless hardening rate and is of the order
$H_h \approx 15$ \cite{quinlan96}. In a fixed isothermal stellar
background with a number density of stars,
$\rho_{\star} = \sigma_c^2/(2~ \pi G r^2)$ the
hardening timescale starting at $a = a_h$ becomes
\begin{eqnarray}\label{eq:binary_hardening}
  t_h &=& \frac{\pi ~G ~M_2}{2 ~\sigma_c^3 ~H_h} \\ \nonumber
  &\approx& 1.44 \times 10^7
  \left(\frac{M_2}{260 ~M_{\odot}}\right) \left(\frac{\sigma_c}{200
  ~{\rm m~s^{-1}}}\right)^{-3} \left(\frac{H_h}{15}\right)^{-1} ~{\rm yr}
\end{eqnarray}
As it stands this implies that the hardening timescale is
proportional to the lighter MBH in the binary system. This makes
sense: as the mass of $M_2$ increases the fraction of energy lost
to any interacting star decreases. A light MBH binary therefore
needs less stellar interactions to harden. However, this is only
strictly valid if the stellar background is fixed. That is, we
have ignored any depletion of the stellar density in the binary
environment due to the interactions and resulting ejections of
stars from the binary (see for example Milosavljevi´c \& Merritt
(2001) \nocite{milosavljevic01}).

The hard binary stage can represent a bottleneck on the way to MBH
mergers -- particularly for very massive (S)MBHs, as dynamical
friction is no longer significant but gravitational radiation not
yet effective enough in further reducing the separation between
the MBHs. However, both, the dynamical interaction with stars in
the MBH binary's vicinity as well as gas infall are likely to have
the potential to quickly reduce the distance between the MBHs
\cite{barnes96,naab01,barnes02,armitage02,escala04}. We do not
quantify this effect, but note that the role of gas infall in this
context is probably more important at early times when MBHs
encounter each other more often in the wake of major mergers of
two host haloes \footnote{At very high redshifts $z \grtsim 10$
the mass ratios $M_2/M_1$ between MBHs are closer to 1 on average
as central SMBHs have not yet grown to the massive sizes observed
today. The more similar mass of MBHs implies that they encounter
each other in the wake of the merger of two haloes that are also
of similar mass.}.

\subsubsection{Dynamical friction at the centre of haloes}\label{sec:df_halocenter}
At still larger distances, dynamical friction is the main mechanism by
which MBHs are drawn to the centre. The dynamical friction force is
given by the Chandrasekhar formula \cite{binney87}
\begin{equation}
  \label{ch5:eq:dynfrict_gen}
  \frac{\d {\bf v}_{sat}}{\d t} = -16 \pi^2 {\rm ln} \Lambda_c G^2 m
  (M_{sat} + m) \frac{\int_0^{v_{sat}} f(v') v'^2 \d v}{v^3_{sat}} {\bf
  v}_{sat}
\end{equation}
where ${\bf v}_{sat}$ is the satellite velocity with respect to the
host centre, $M_{sat}$ the satellite mass, $m$ is the mass of individual
host halo particles, that have a distribution function $f(v)$, and
ln $\Lambda_c = {\rm ln}(b_{max}/b_{min})$ is the Coulomb
logarithm, that is the ratio of maximum and minimum impact parameters,
$b_{max}$ and $b_{min}$,
of the host particles with respect to the satellite system.

This effect has been modelled in
the numerical scheme using the semi-analytical dynamical friction timescale given by
Colpi, Mayer \& Governato (1999; see also Taylor \& Babul 2004) \nocite{colpi99,taylor04}
\begin{equation}\label{eq:dynfrict_colpi}
  \tau_{df} = k \frac{(M_{host}/M_{sat})}{\ln (M_{host}/M_{sat})}\epsilon^{0.4}
  \frac{P_{vir}}{2~\pi}
\end{equation}
where $M_{host}$ and $M_{sat}$ denote the mass of the host and satellite haloes
respectively and $P_{vir}$ is the circular orbital period at the virial radius
of the host, and $k$ is a numerical constant.
Colpi \ea(1999) find $k = 1.2 e \simeq
3.26$ for massive satellites in an isothermal potential with a
constant-density core. We will use the value k = 2.4, found to match
the infall times for massive satellites in a cuspy potential in the
model of Taylor and Babul (2004).
The circularity parameter $\epsilon = J(E)/J_{circ}(E)$
is the ratio of the angular momentum of the actual satellite orbit and that
of a circular orbit with the same energy.

In the following we assume that this
timescale dominates both the timescale for gravitational wave
induced coalescence and that for overcoming the hard binary stage,
which in principle can be very short due to the effects of gas and
stellar dynamical processes.

In order to compute the time when a MBH merges with the central (S)MBH
in the host we first establish the time when a satellite/MBH system sinks to an
orbit with a radius less than the radius of the infall region. This is
1 per cent of the host virial radius at the time. At this point the
satellite/MBH system in question has so far been  considered as having `fallen to the
centre'; following their dynamics to smaller radii was too
expensive computationally.

\begin{figure}
  \includegraphics[width=8.0cm]{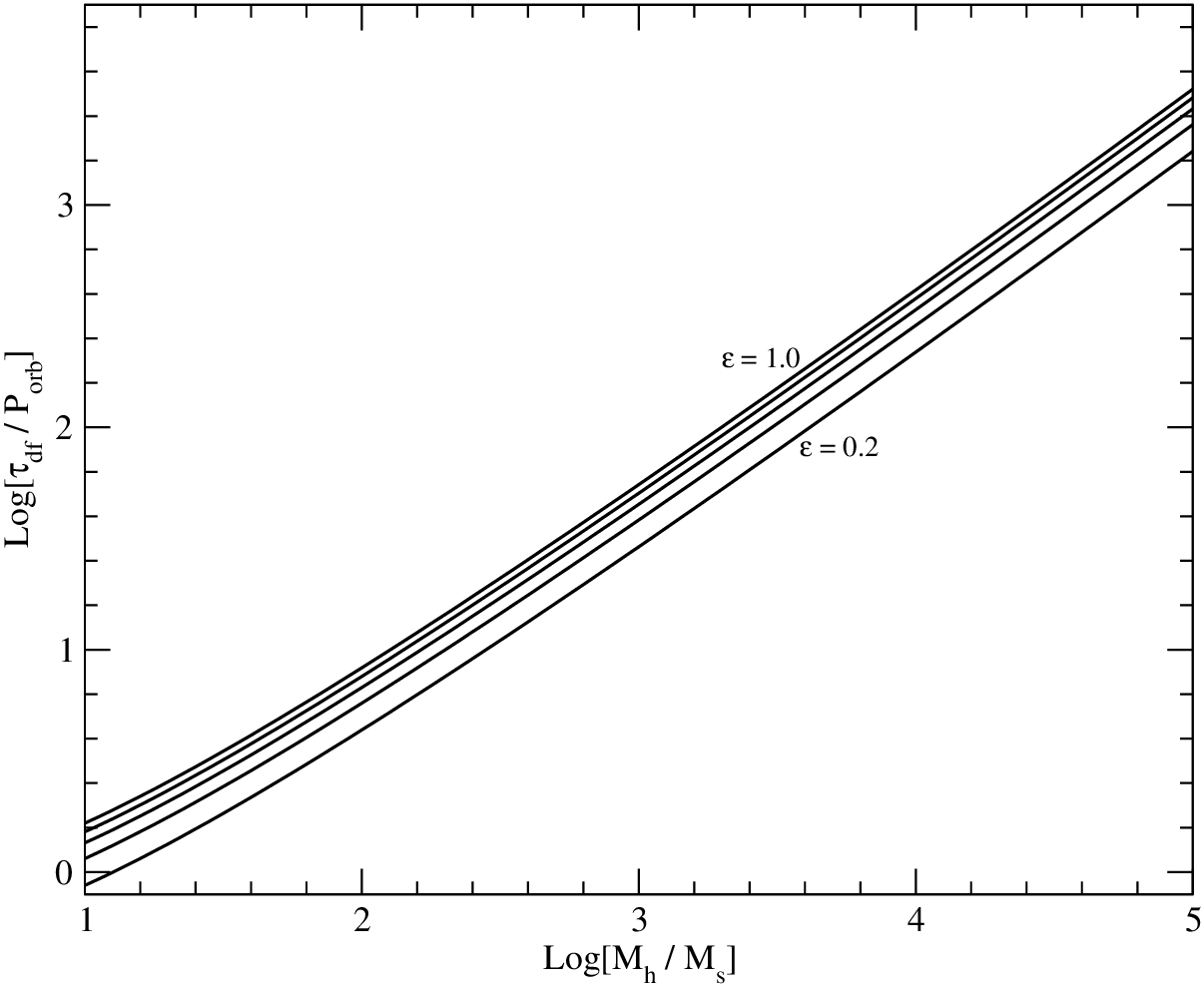}
  \caption{
    Ratio of dynamical friction time to orbital period plotted
    against ratio of host mass interior to satellite orbit to satellite
    mass. From top to bottom the curves are for values of the
    circularity parameter $\epsilon = 1.0, 0.8, 0.6, 0.4, 0.2$.}
  \label{fig:colpi_dynfrict}
\end{figure}

To track infall more accurately, we add to this time
a dynamical friction infall time, using
eq.(\ref{eq:dynfrict_colpi}) and replacing the total host mass
interior to the virial radius and the orbital period at the
virial radius with the
respective quantities at the outer radius of the infall region. This
is also shown in figure \ref{fig:colpi_dynfrict} for a range of
satellite to
host mass ratios. The circular parameter $\epsilon$ is computed from
the last orbit of the satellite system before it crossed into the
infall region.
The result is a more accurate estimate of the time when the actual
merger occurs. As a result of the correction to the infall time we
also inevitably change the order in which the MBHs merge with the
centre. This in turn leads to a change in the growth history of the
central (S)MBH  and also to a slightly lower final mass of the
SMBH, since there is now a small fraction of satellite MBH systems
that do not merge with the centre within a Hubble time.
\begin{figure}
  \includegraphics[width=8.0cm]{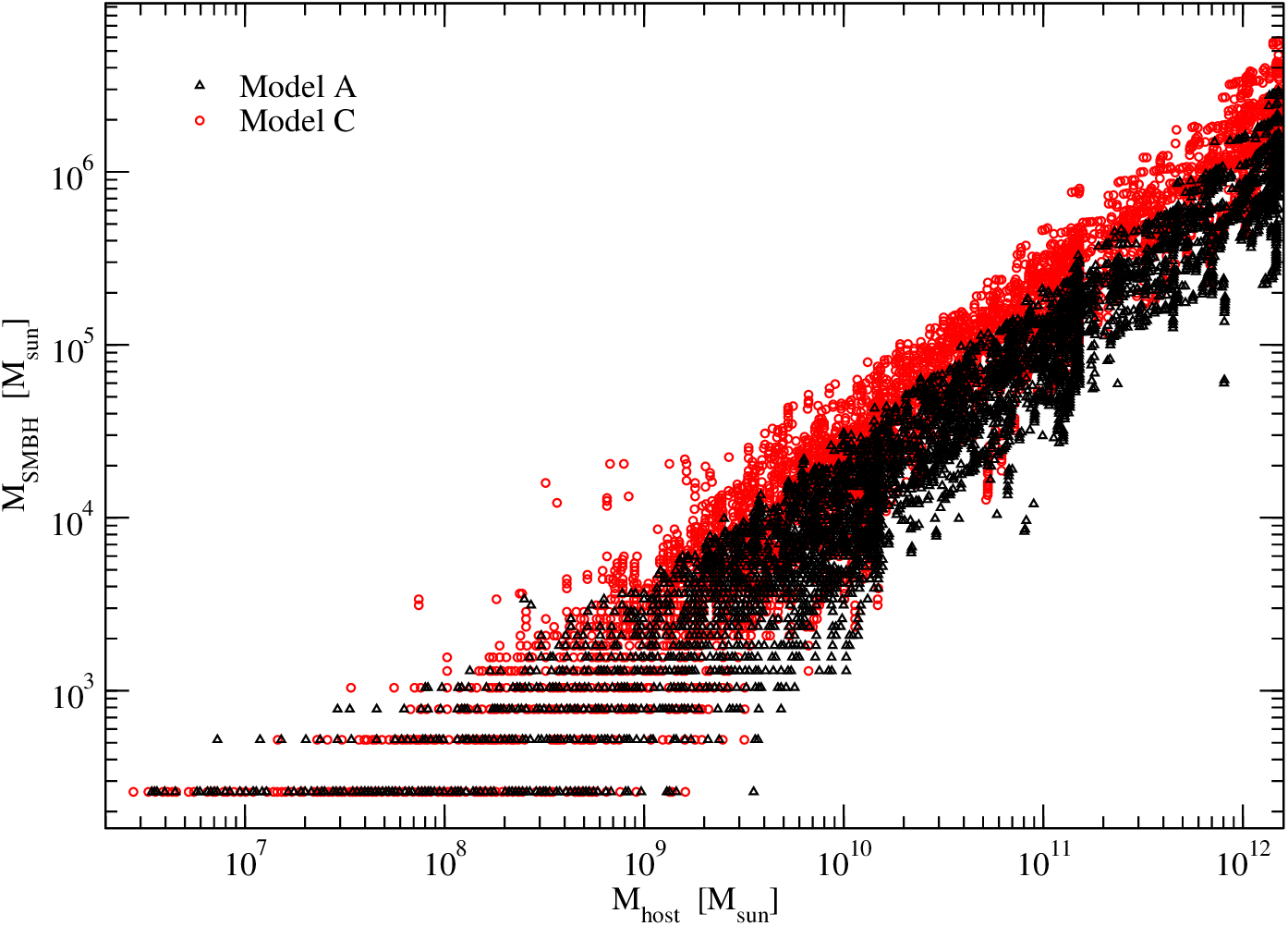}
  \caption{The figure shows the total mass of the host halo versus
    the mass of the
    central (S)MBH. For the (S)MBH mass we have taken account of the
    slightly altered growth history due to the addition of a
    dynamical friction infall time after satellite MBH systems
    have crossed the infall radius.
  }
  \label{fig:Mhost_Msmbh}
\end{figure}
Figure \ref{fig:Mhost_Msmbh} shows the host halo mass versus the
resulting mass of the central (S)MBH for final halo masses of $1.6
\times10^{10,11,12}$ \Msun\ and their lighter precursor haloes at
higher redshifts\footnote{The SMBH to halo mass has also been
determined on the basis of velocity dispersion data outside the
bulge dominated regions \cite{ferrarese02}}. The data points from
precursors of different final halo masses are recognisable by
their accumulation near those final halo masses. Nevertheless, all
data points lie on a line with a slope that is also consistent
with the one we determined for the $M_{SMBH}-M_{bulge}$ relation
in paper I. This indicates that the ratio of SMBH to halo mass
does not depend significantly on the redshift.

We also made an important assumption regarding the actual mass of the
satellite systems in which the MBHs are embedded. The semi-analytical
code we used to follow the satellite dynamics considers every
satellite naked if its mass has dropped to something less than about
0.3 per cent of its original mass \footnote{The actual criterion
  refers to a limiting tidal radius of a satellite system, beyond
  which it is considered `naked'. This criterion roughly corresponds to
  the 0.3 per cent mass limit we used.}.
For all `naked' MBH satellite systems arriving at the infall radius we
have therefore assumed a mass equal to this limit. We do impose a
lower limit of $3\times10^3$\Msun\ for the mass of any naked satellite
system, roughly corresponding to 1 per cent of the mass of the minihaloes
and large enough for the presence of a baryonic core of the order
the MBH mass.
It is this remnant material of the satellite associated with an MBH
that allows MBHs to the delivered to the centre efficiently by
dynamical friction. This point has also been mentioned by Yu (2002)
\nocite{yu02a}.

For the halo of final mass $1.6 \times 10^{12}$ \Msun\ we find the
following.
Of the satellite MBH systems crossing the infall radius
about 96 per cent do arrive at the centre within a Hubble time. These MBHs
contain 99 per cent of the mass of all MBHs in the infall region. If we
reduce the satellite mass to 0.03 per cent of its original value and the
lower limit to $ 3\times10^3$\Msun\ we still find that 88 per cent of the
MBHs in the infall region (accounting for
97 per cent of the mass) end up at the centre within a Hubble time.
In haloes of lower final mass virtually all MBHs systems crossing the
infall radius travel to the centre within a Hubble time.

\subsubsection{Dynamical Friction and Gas Infall in Galaxy Mergers}
MBHs with masses and orbits that would not allow them to travel to the
centre within a Hubble time may still end up in the centre if the
environment changes in a way that allows dynamical friction to act more efficiently.
Mergers between galaxies of similar mass (`major mergers') through the
accompanying violent dynamical processes may induce the dramatic
changes to the matter distribution in the central region that we are
looking for. Hydrodynamical simulations of mergers of gas rich
galaxies \cite{barnes02,barnes96,naab01} show that up to $60$ percent
of the total gas mass of the two galaxies can end up within a region
just a few hundred pc across at the centre of the merger remnant,
thereby triggering starbursts and initiating the fuelling of
quasars\footnote{The galaxy interaction leads to the shock heating
  of the gas and the formation of a bar in the gas
  distribution. Radiatively cooled gas then falls towards the centre
  along the bar.}.
This highly dense gaseous core should remain until feedback from
star formation, supernovae or quasar activity begins to drive it out again.

We have already identified gas infall as a potential mechanism for
accelerating the evolution
of a hard MBH binary. We now look at how this effect can also boost
the efficiency of dynamical friction as MBHs move in the central region
of the merger remnant.
If the respective central-kpc region of each galaxy ends up within a
kpc or so of the new centre of the remnant after merging, then so will
any MBHs that were present within a kpc of the original galactic centres.
This seems reasonable since the central regions, sitting deepest in
the gravitational potential of each galaxy, will be most stable to
tidal disruption. It is plausible that MBHs originally present there
travel along the line connecting the centres of the merging galaxies, towards the new
global potential minimum of the merger remnant.

This migration of MBHs could be accelerated by the rapid infall of large amounts of gas into a
small core region. The question then is whether the
density and the bulk inflow velocity of the gas is sufficient to
accelerate any orbiting MBH towards the centre via dynamical friction.
The complex dynamical events and rapidly varying geometry make
it difficult to determine accurately any radial bulk inflow velocities
and gas densities as they evolve with time, and particularly so in the
core central region, where things are made more difficult by the
resolution limits of the simulations.
Even a rough estimate of these quantities may already yield
useful constraints for our problem, however.
In the following we try to estimate the timescale for infalling gas
to carry two MBHs towards the centre by dynamical
friction before the gas leaves the central region again.

For two disk galaxies, each having a total mass $M_{gal} = 2.75 \times 10^{11}$
\Msun\ and gas mass $M_{gas} = 5.5 \times 10^9$ \Msun\ in the disk
and initial pericentric separation of $R_p = 8 {\rm kpc}$, 
Barnes \& Hernquist (1996) find
that roughly $60$ percent of the total gas mass end up within a region
approximately $100$ pc in radius some 1.5 Gyrs later.
Assuming the centre of the merger remnant will be half-way between the
galaxies at initial separation and the gas flows towards this
point along the line connecting the merging galaxy centres, the
average inflow velocity of the gas is $v_{flow} \sim 4 {\rm
kpc}/1.5 {\rm Gyr} \approx 2.7 ~{\rm km s}^{-1}$. Further we assume
that the gas density in the flow is of the same order as the final gas
density in the centre, i.e. $\rho_{flow} \sim 7\times10^9 M_{\odot}/(4
\times 100^3 {\rm pc^3}) \approx 1.6\times 10^3 {\rm Mpc ~pc^{-3}}$.
Putting this into eq. \ref{ch5:eq:dynfrict_gen} we find that the MBHs
get accelerated to within 1 per cent of $v_{flow}$ in a time
\begin{equation}
  \tau_{df,flow} \sim 2\times 10^6 \left(\frac{M_{\bh}}{260
  ~M_{\odot}}\right)^{-1} ~{\rm yr}
\end{equation}
Even for seed mass MBHs, this time is negligible compared to the time
scale of the inflow of about 1--2 Gyrs.
Although we do not know what exactly will happen to the MBHs once they have
entered the central 100 pc of the remnant, gas inflow seems at least capable of
efficiently transporting MBHs from the outer radius of the infall
region ($\sim 1 {\rm kpc}$, for galaxies of this size) down to at
least 100 pc. This holds in particular for MBHs that previously would not
have been able to spiral anywhere near the centre.

\subsubsection{Multiple MBH interactions}
Up to now we have implicitly assumed that a MBH binary merges
before it forms a new binary with another incoming MBH. However,
given the number of inspiralling MBHs involved it is possible that
there are cases where incoming MBHs will find another MBH binary
that has not yet merged. If the binary system and the incoming
third MBH come close enough then there is a possibility that one
of the MBHs will be ejected \cite{begelman80}. The required
proximity of an incoming MBH and MBH binary is facilitated primarily
by major mergers. This will be the case particularly at high
redshifts, as most MBH interactions will be between MBHs with
similar masses (of the order of the MBH seed mass), which also
implies similar masses of their associated haloes.

As far as our model is concerned we have not incorporated the
possibility of triple interactions in our results. We assume that
stellar and gas dynamical effects in the wake of major mergers always
lead to swift merging of MBH binaries, primarily through accelerating
evolution of the hard binary stage.
Here we consider what happens if we drop this assumption.
For the case that stellar and gas dynamics have no mitigating effect,
we can determine a rough limit for the number of triple
interactions and resulting sling-shot ejections of MBHs.
\begin{figure*}[t]
  \hbox{\resizebox{\hsize}{!}{\includegraphics{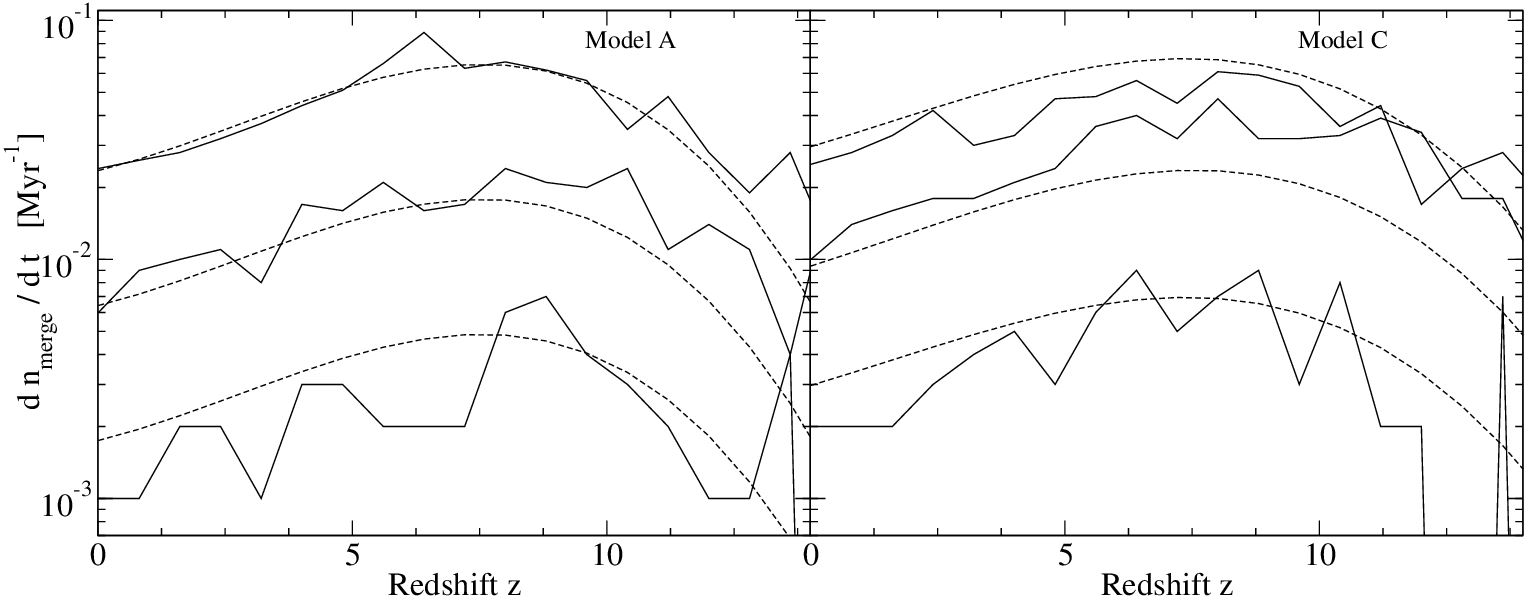}}
  }
  \caption{
    Rates of halo MBHs merging with central (S)MBH,
     as a function of redshift. Results are shown
    for models A and C and for final halo masses of $1.6 \times
    10^{12,11,10}$\Msun\ (three sets of curves, from top to bottom).
    The dashed curves
    represent the same best fits to the data scaled by a factor
    corresponding to the different final halo mass (see text).
      }
  \label{fig:mergrate_individhaloes}
\end{figure*}

In figure \ref{fig:mergrate_individhaloes} we show the number of
mergers per unit time ($10^6$ yr) for model A and C. From top to
bottom the curves are for $1.6 \times 10^{12,11,10}$\Msun\ haloes. The
dashed lines are the best fits.
Let us consider the largest merger rate, which we find at $z \sim 5 -
10$ in the precursors of haloes with final mass
$1.6\times10^{12}$\Msun.
The rate is $\d N_{merge} /\d t \sim 0.07 {\rm ~Myr}^{-1}$ implying
that two MBHs arrive at the centre within $\sim 15$ Myrs.  For the
lowest rates $\d N_{merge} /\d t \sim 0.001 {\rm ~Myr}^{-1}$ two MBHs
would reach the centre within about a Gyr. These timescales would
have to be compared with the hardening time of a MBH binary at
the centre to determine the likelihood of a triple interaction.
Yu (2002) \nocite{yu02a} finds that the hardening timescale can be
significantly larger for equal mass binary systems, which again is
more likely to be the case at very high redshifts.

Although the central halo MBH merger rate as shown in figure
\ref{fig:mergrate_individhaloes} is highly uncertain at high
redshifts, there is a clear declining trend towards high
redshifts. Consequently we would expect multiple MBH interactions to
be less important. Overall, this is different from Volonteri ~\ea (2003)
\nocite{volonteri03}, who find that there is a very significant
probability for triple interactions. One important reason for this
discrepancy is that they explicitly did not take into account tidal
stripping of satellite systems and the resulting increase in the
dynamical friction infall timescale.
When MBHs are involved in triple interactions, they find that
those MBHs that get ejected MBH will actually leave the
galaxy (remnant) altogether \cite{volonteri03}. These ejected MBHs
could thus constitute a population of inter-galactic MBHs. Due to the
nature of the process these MBHs will be mostly seed mass MBHs and
will have been stripped naked, so that they will be virtually
impossible to detect in the IGM.

The overall result is then, that while the formation of multiple MBH
systems is possible, the probability of such events is likely less
important than analytical arguments suggest.
Previous estimates typically did not account for the tidal stripping
of infalling satellites let alone their detailed dynamical evolution,
but also the role of gas dynamics at the centre. The first effect
implies that the rate at which MBHs arrive from the outer parts of the
halo at the halo centre is lower, whereas the latter means that once
MBHs have arrived at the centre, they form a binary that coalesces faster.
In the light of this, we believe our assumption of efficient merging
to be justified, particularly at redshifts $\lesssim 15$ where MBHs
mergers actually produce a strong enough gravitational wave signature
that could be detected as we will see in the next section.


\subsection{Rate of SMBH-MBH mergers}
\begin{figure*}[t]
  \hbox{\resizebox{\hsize}{!}{\includegraphics{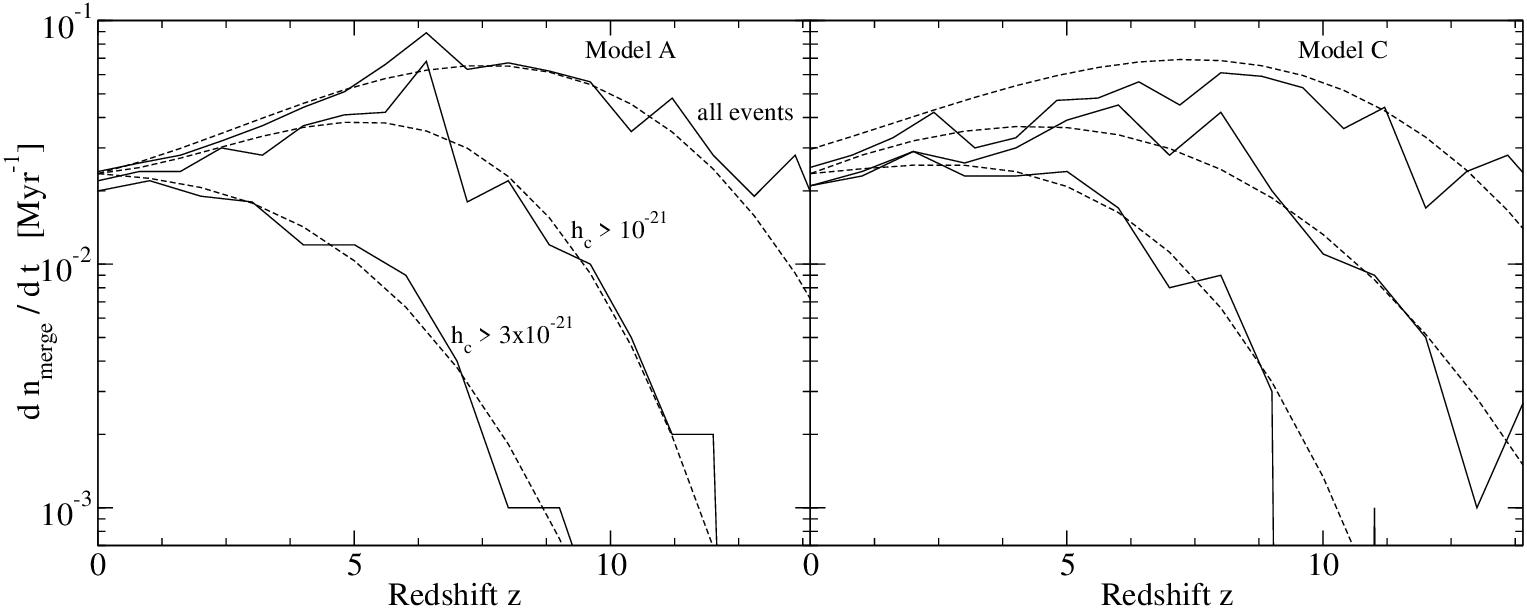}}
  }
  \caption{
    Rates of halo MBHs merging with central (S)MBH,
    as a function of redshift. Results are shown
    for models A and C and final halo mass of $1.6 \times
    10^{12}$\Msun\ with the dashed line representing a best fit. From
    top to bottom, the three sets of curves show the contribution
    of events with dimensionless strain amplitudes above a given value
    as shown.
  }
  \label{fig:mergrate_16e12halo_hstrain}
\end{figure*}
We now attempt to obtain the cumulative signal of all mergers. To do
this we essentially need to integrate over the mass function of
haloes times the number of mergers per halo and finally integrate again
over the relevant redshift ranges.

In what follows, we will only consider models A and C. In model D the
rate of mergers is too low to derive a meaningful estimate of the
average merger rate per halo, and consequently the total rate of events
received from across the sky. Model B only
differs from A in that it has a larger seed MBH mass. We do not expect
this to make a significant difference when it comes to mergers with
the centre. When calculating the dynamical friction time from the
point the MBH systems cross the infall radius (cf.~section
\ref{sec:df_halocenter}), we assumed the MBH satellite system's mass to
be $\sim 0.3$ per cent of the original satellite mass. The difference in the
satellite's MBH mass in model A and C is thus insignificant. The
larger MBH mass in model B will also only lead to very slightly
increased strain amplitudes when MBHs eventually merge.

The number of MBH mergers per halo is shown in figure
\ref{fig:mergrate_individhaloes} for models A and C. The dashed lines
represent a single  cubic-polynomial fit in log space that is
multiplied with a simple scaling factor for different halo masses.
\begin{eqnarray}
  \frac{\d n_{merge}(M_{h,0}, z)}{\d t} &=& \left(\frac{M_{h,0}}{1.6\times
  10^{12} M_{\odot}}\right)^\beta \nonumber \\
  & \times & e^{{\rm b}_0 + {\rm b}_1~z + {\rm
  b}_2~z^2 + {\rm b}_3~z^3} ~{\rm Myr}^{-1}
\end{eqnarray}
where $M_{h,0}$ denotes final halo mass.
For model A we find $({\rm b}_0,{\rm b}_1,{\rm b}_2, {\rm b}_3,\beta)
= (-3.75, 0.12, 0.022, -2.63\times10^{-3}, \frac{5}{9})$, while for model C
we find $(-3.52,
0.145, 7.5\times10^{-3}, -1.57\times10^{-3}, \frac{1}{2})$.
In this form the merger rate can now simply be divided into a product
of a redshift dependent part, $u(z)$, and another one scaling with final halo
mass
\begin{equation}\label{eq:halomergrate}
  \frac{\d n_{merge}}{\d t}(M_{h,0}, z) = \left(\frac{M_{h,0}}{1.6\times
  10^{12} M_{\odot}}\right)^\beta ~u(z)
\end{equation}

Figure \ref{fig:mergrate_16e12halo_hstrain} shows the merger rate per
halo with strain amplitudes $h_c > 10^{-21}$ and $3\times10^{-21}$ for
a final halo mass of $1.6\times10^{12}$ \Msun\ only. While the best
fit parameters change accordingly, we have assumed that the scaling
with halo mass, as parametrised by $\beta$, remains the same. Although
we have not explicitly stated any error bars, uncertainties will be
quite large, since the merger rate per halo is determined by averaging
over individual merger events occurring over large timescales. The
only way of reducing this uncertainty is to run a significantly larger
number of merger trees.

When determining $h_c$, we assume that the merging MBHs emit waves
at the maximum frequency which is only expected to occur very briefly
towards the end of the inspiral phase due to the rapid sweeping up in
frequency.
This, however, should not affect our result significantly, since $h_c$
depends only weakly on the frequency (cf.~equation (\ref{eq:hc})).

Using the merger rate per halo we can already determine as a function of epoch a lower
limit to the rate, ${\rm d}
\nu_{rec}$ of received SMBH-MBH merger events.
\begin{eqnarray}\label{eq:densityevents}
    \frac{{\rm d} \nu_{rec}(z)}{{\rm d} V} & >&  \frac{1}{1+z}
    \nonumber \\
    & & \times \int_{M_{h,min}}^{\infty} \frac{{\rm d}
n_{h}}{{\rm d} M_{h,0}} \frac{\d n_{merge}}{\d t}(M_{h,0}, z)
    ~{\rm d} M_{h,0} \nonumber \\
    & = & \frac{u(z)}{1+z} \nonumber \\
    & & \times \int_{M_{h,min}}^{\infty} \frac{{\rm d}
n_{h,0}}{{\rm d} M_{h,0}}  \left(\frac{M_{h,0}}{1.6\times
  10^{12} M_{\odot}}\right)^\beta ~{\rm d} M_{h,0}\nonumber \\
\end{eqnarray}
where $V$ denotes volume and (1 + z) accounts for the redshift induced
reduction in the rate of received events. ${\rm d} n_h/{\rm d}
M_{h,0}$ is the differential mass function of haloes of mass $M_{h,0}$
\begin{equation}\nonumber
\frac{{\rm d}n_{h}}{{\rm d}M_{h,0}} = \sqrt{\frac{2}{\pi}}
\frac{\rho_0}{M_{h,0}} \left|\frac{{\rm d}{\rm ln}\sigma}{{\rm d}{\rm
~ln}M_{h,0}} \right| \frac{\delta_c}{\sigma(M_{h,0})} {\rm exp}\left[\frac{-
\delta_c^2}{2 \sigma(M_{h,0})^2}\right]
\end{equation}
where $\rho_0$ is the present-day cosmic matter density, $\sigma(M_{h,0})$
is the variance in the linear matter density field
on scales corresponding to mass $M_{h,0}$,
 and $\delta_c$ is the critical overdensity for collapse.

Equation \ref{eq:densityevents} is a lower limit, as it only considers
mergers in the main trunk of the tree that grows to become a single
halo of mass $M_h$ at $z = 0$. To account for the mergers in side
branches before they become incorporated in the main trunk of the
tree the Press-Schechter (PS) mass function, which we had only
computed for final halo masses above, needs to be replaced by the
redshift dependent mass function.
In addition we need to re-express the redshift dependent MBH merger
rate per halo in terms of the halo mass at the redshift concerned.
From our simulations we found that the average mass of a halo at
redshift $z$, given that it grows into a halo of final mass $M_{h,0}$
today, can be approximated by
\begin{equation}\label{eq:halogrowth}
  M_h(z) = M_h(M_{h,0},z) \approx \Exp\left[12 + \tilde{m} - 0.04
  ~\tilde{m} ~z\right]
  \end{equation}
where $\tilde{m} = 11.5 +
\ln\left(\frac{M_{h,0}}{1.6\times10^{10}M_{\odot}}\right)$.
We can now substitute this into equation \ref{eq:halomergrate} to
eliminate the dependence on $M_{h,0}$ and
obtain
\begin{equation}
   \frac{\d n_{merge}}{\d t}(M_h, z) =
   \left[10^{-7} \times \left(\frac{M_h}{1.6\times10^5
   M_{\odot}}\right)^{\frac{1}{1 - 0.04 z}}\right]^\beta ~u(z)
\end{equation}

\begin{figure*}[t]
  \hbox{\resizebox{\hsize}{!}{\includegraphics{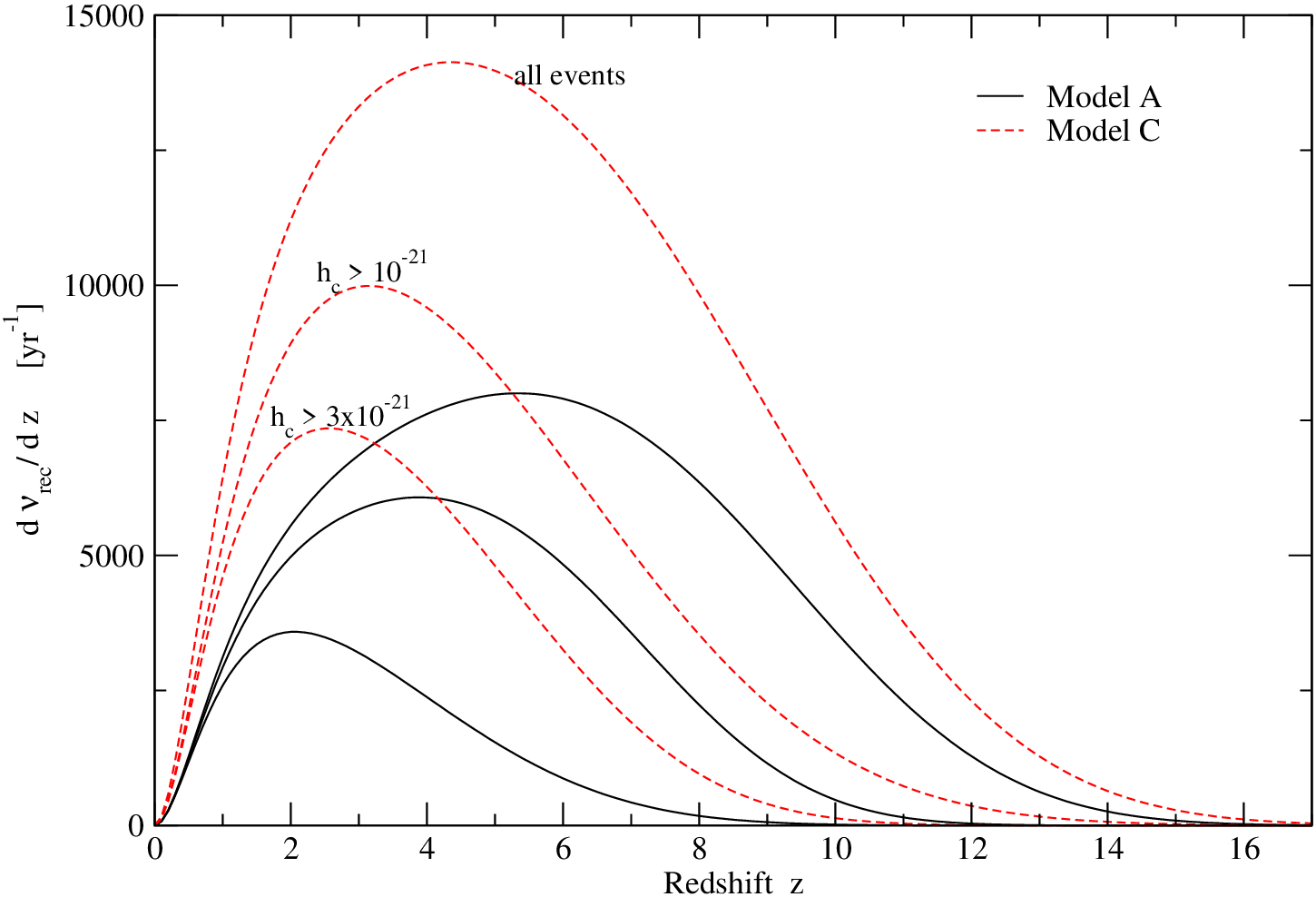}}
  }
  \caption{Redshift dependent rate of received merger events for models A and
    C. From
    top to bottom, the three sets show the contribution
    from all events and those with dimensionless strain amplitudes
    above $10^{-21}$ and $3\times10^{-21}$ respectively.
  }
  \label{fig:received_differential_rate_Gwaves}
\end{figure*}

The lower integration limit in eq \ref{eq:densityevents} is
determined by the minimum mass of haloes that harbour
central MBHs. This mass has to be at least of order $10^7$
\Msun, since we argued above that only at this mass has every halo
acquired one seed MBH. This is important since we have implicitly
identified the number density of haloes with that of central MBHs.

If we now multiply $\frac{{\rm d} \nu_{rec}(z)}{{\rm d} V}$ with the
volume of a spherical shell at radius $D(z)$ we
obtain the total rate of events, d $\nu_{rec}(z)$, received from sources at distance
$D(z)$ to $D(z) + {\rm d} D$
\begin{equation}
    {\rm d} \nu_{rec}(z) = \frac{{\rm d} \nu_{rec}(z)}{{\rm d} V} 4 \pi D(z)^2 {\rm d}D
\end{equation}
The resulting differential redshift distribution of the received rate
of events is shown in figure
\ref{fig:received_differential_rate_Gwaves}, where we have also shown the
contribution from events with associated dimensionless strain
amplitudes $h_c$ above $10^{-21}$ and $3\times10^{-21}$. The latter is
based on the fits to the merger rate per halo shown in figure
\ref{fig:mergrate_16e12halo_hstrain}.
Again, the errors are quite large and are
difficult to quantify precisely, primarily because of the
uncertainties in the merger rate per halo to which we now also have to
add the uncertainty in the halo mass scaling, which we described by a
power law with index $\beta$.

This differential distribution can then be integrated over all
redshifts to obtain the total number of merger events received.

\subsection{Detections and the distribution of strain amplitudes} \label{sec:LISAlimit}
For actual detections with LISA, we need to take into account its
restrictions on the minimum detectable strain amplitudes. For the MBH
masses under consideration here, and the resulting wave frequencies, LISA
should be able to see events with $h_c \sim 10^{-21}$, and may even
see events as weak as $h_c \sim 10^{-23}$ near $f_{gw} \sim 0.01$ Hz.
For the combination of MBH masses and the redshift range considered,
all events yield strain amplitudes larger than $10^{-23}$, and LISA
may therefore be able to detect most if not all of these events.
Adopting the more conservative LISA limit of $h_c \sim 10^{-21}$ the
number of detections is somewhat smaller as shown by the
corresponding curve in figure \ref{fig:received_differential_rate_Gwaves}.

We are interested in detections of events at
cosmological distances, and to do so the conservative LISA detection
limit requires a minimum chirp mass of
about $10^4$ \Msun\ \cite{haehnelt94}. MBHs of this size, however, will typically be
hosted inside haloes of at least $10^9$ \Msun, which could therefore be taken as the
effective lower limit of integration in equation \ref{eq:densityevents}. In fact ${\rm d}
\nu_{rec}$ is not very sensitive to the particular choice of
integration limits. The event rate of mergers in haloes as determined
by our simulations is very small at the low-mass end, which therefore
does not contribute significantly to the overall rate. Similarly, the
steep decline in the mass function at high masses imposes an effective
upper limit.

\begin{figure*}[t]
  \hbox{\resizebox{\hsize}{!}{\includegraphics{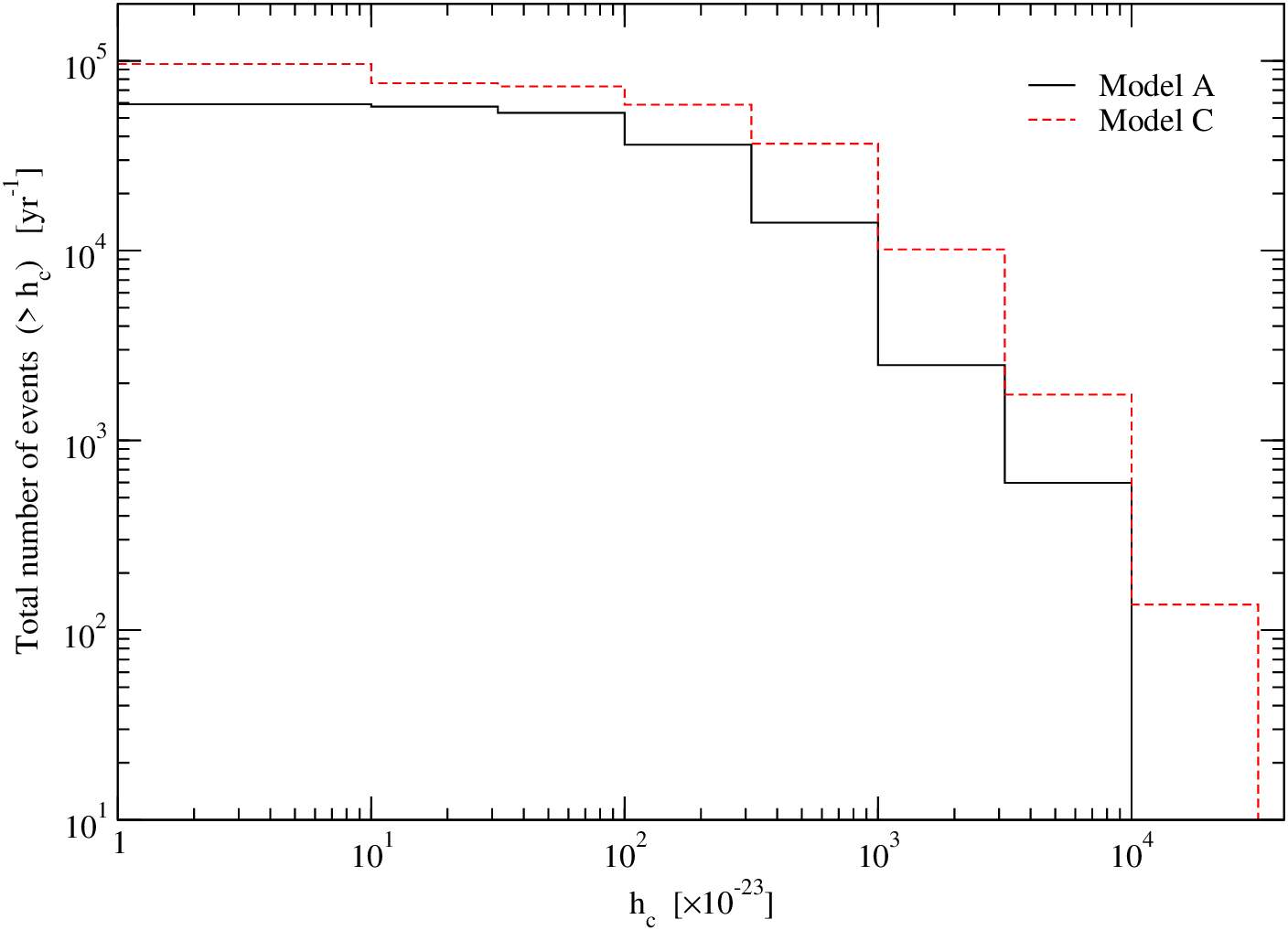}}
  }
  \caption{
    Distribution of dimensionless strain amplitudes for events
  received from across all redshifts.
  }
  \label{fig:Gwave_strain_distrib}
\end{figure*}

The redshift distribution of merger events can now be integrated
across the redshift range to obtain the total number of events
received.
The result is shown in figure \ref{fig:Gwave_strain_distrib}, where
we have shown the total number of events above a given strain
amplitude. As in the previous section, we have used a fit to the merger rate
per halo above a given strain amplitude to determine the merger rate
per unit redshift.
From this we see that LISA would be expected to detect some $10^4 -
10^5$ events per year. The difference between model A and C
is too small to be detectable in practice, as the overall
uncertainties in our results are of at least the same order.

We have seen above that the strain amplitude is rather insensitive to
the frequency of the waves emitted. The frequency does matter for LISA
detections as its peak sensitivity lies in a rather narrow band around
0.01 Hz. Merging seed MBH binary systems have very low strain amplitudes,
especially at high redshifts, but peak frequencies well above 1
Hz. The only way of detecting these is during a significantly earlier
phase of their coalescence when their frequency is lower.

\section{Gamma rays from MBH formation and merging}
\label{sec:GRBs}
Another exciting possibility is that the more massive of the first
stars forming at high redshift conclude their lives in hypernovae,
which be visible as gamma-ray bursts (GRBs). GRBs should be detectable
to high redshift by upcoming satellite missions, which would allow
us to study events from the first era of star formation directly.
In contrast to the other observations mentioned above, such observations
cannot be used to verify the hierarchical MBH merging scenario.
GRB observations might provide constraints on the abundance of the first
massive stars, however, and in this way allow us to test this
part of our model directly.

GRBs are the brightest explosive events in the
Universe. Typically they involve the emission of large amounts of
radiative energy, of the order $\sim 10^{52 - 54}$ ergs in $\gamma$
rays, over a period of only a few seconds.
A GRB is followed by an afterglow in the X-ray, optical or radio
parts of the spectrum and lasts from days to
weeks. The total energy emitted in an afterglow is typically two or
three orders of magnitude below that of the actual burst.
Spectroscopy has been carried out on the afterglow emissions as well
as the emission of what appear to be host galaxies of the GRBs,
 once the afterglow
itself has faded away sufficiently. In this way it has been possible
to constrain the
redshift of a number of GRBs below $z < 3$. If GRBs could be used as
standard candles in a similar way as type Ia supernovae, this would facilitate
their use to calibrate the GRB distance-luminosity relation. On this
basis one could then determine the maximum redshifts at which GRBs
could still be detected by present and future observing missions, such
as with the BATSE detector on the {\it Compton Gamma-Ray
  Observatory} or the {\it SWIFT} satellite due to be launched late in 2003
\cite{lamb00}. For a number of GRBs for which redshifts could be
established, Lamb \& Reichart (2000) found that {\it SWIFT} could have
in principle detected these out to redshifts of $z \sim 20$, and some
to $z \sim 70$.
For further details of GRBs their afterglows the reader is referred to the review by Piran (2000)
\nocite{piran00} and the extensive work by Bloom (2002)
\nocite{bloom02} and references therein.
For the following we assume that GRBs are in principle detectable out
to redshifts beyond $z > 20$.

A range of processes have been  discussed that could be the cause for
GRBs. As mentioned, one of these is the possibility that GRBs are
produced by the
collapse of massive stars.
In the following we exclusively focus on this possibility and attempt
to estimate the number of GRBs from massive population III stars in
our model.

\subsection{Number of GRBs from massive star collapse}
The procedure to compute the total number of GRBs across all relevant
redshifts is very similar to the one we used to determine the
gravitational wave events. It is different in that we only deal with
the very first stars, and are not concerned with
the subsequent merging of the MBHs these stars leave behind. This is
therefore a purely analytical argument with no recourse to our
simulation data concerning the merging and dynamical evolution of MBHs.

Again, we estimate the abundance of minihaloes at the redshifts in
question ($z \sim 25$). Depending on how many massive stars form
per halo, we can then determine the expected number of GRBs. A
similar route has been followed in previous work. In this case the
abundance of massive stars per halo/galaxy was assumed to trace
the star formation rate at the respective redshift
\cite{bromm02a,choudhury02}. We assume that every minihalo between
a maximum redshift $z_{max}$, corresponding to a
$2.5$--$3.5\sigma$ peak, and a minimum redshift determined by the
end of first star formation, which we take as $z_{min} \sim 20$,
produces one star that explodes as a hypernova with an associated
GRB.

We have seen that a hypernova leads to the expulsion of all gas
from minihaloes. Each minihalo can thus only ever produce one GRB.
This is a problem as we now require each halo to retain a memory
of whether and when it produced a GRB. For this reason it is not
strictly correct to use the Press-Schechter mass function to
compute the number density of haloes at a given redshift and
equate the result with the number density of GRBs. A halo more
massive than $M_{min}$ may have already lost all its gas through a
hypernova at some earlier stage when it was significantly less
massive. In practice, however, the Press-Schechter approach is
more than adequate for two reasons. First, the PS mass function is
steepest for large masses corresponding to haloes collapsing from
high peaks as is the case for all minihaloes in our redshift
range. The number density of haloes with mass $M_h \geq M_{min}$
is therefore vastly dominated by haloes with masses very close to
$M_{min}$. Secondly, the survival time \footnote{ The survival
time is the time by which the mass of a halo has more than
doubled.} for haloes collapsing from high peaks is very short
\cite{lacey93}. That means that by far most haloes with $M_h \geq
M_{min}$ will have formed at or very shortly before the redshift
in question and the probability of previous star formation and a
GRB is very small.

\begin{figure*}
  \vbox{\includegraphics[width=8.5cm]{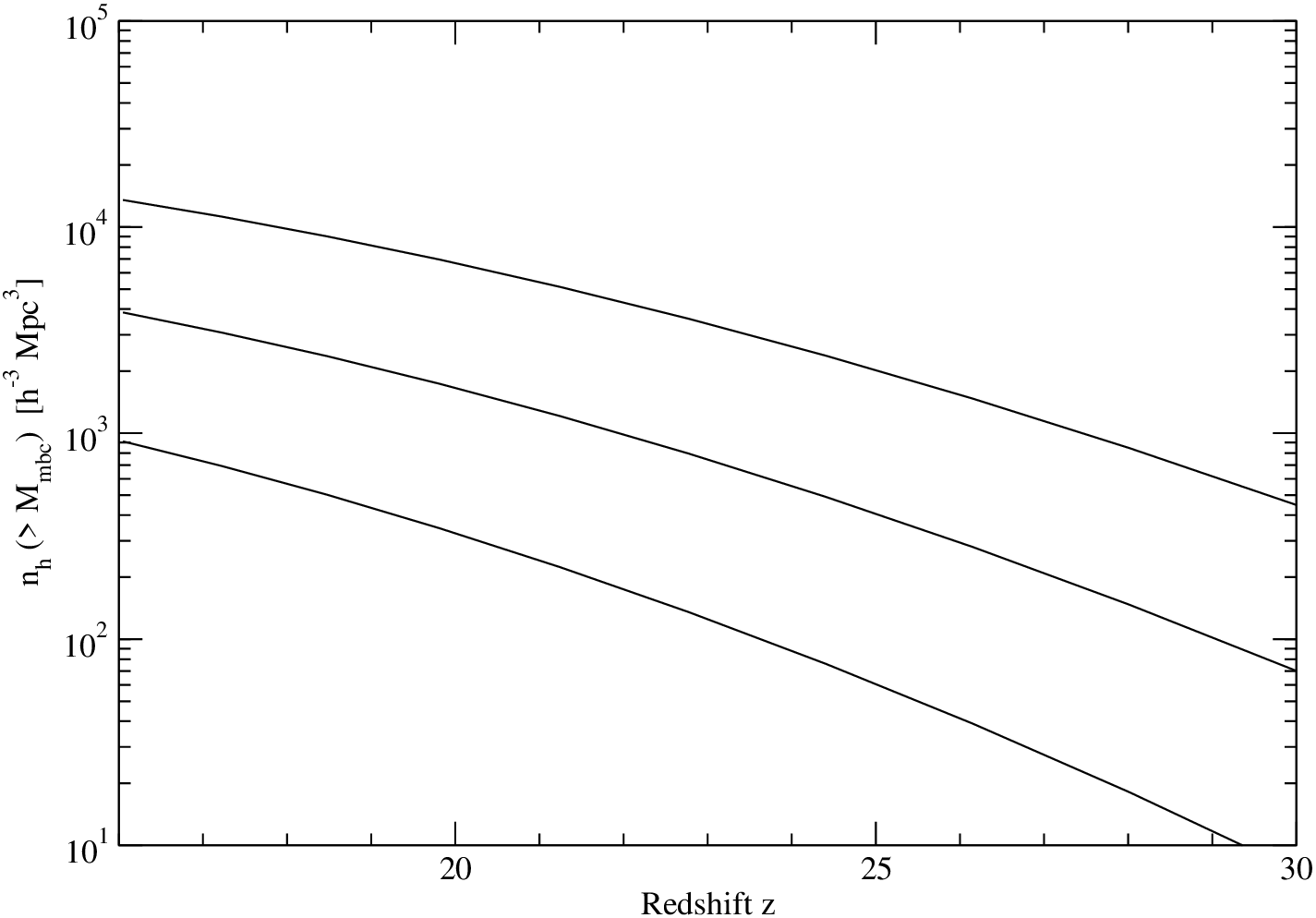}
    \includegraphics[width=8.5cm]{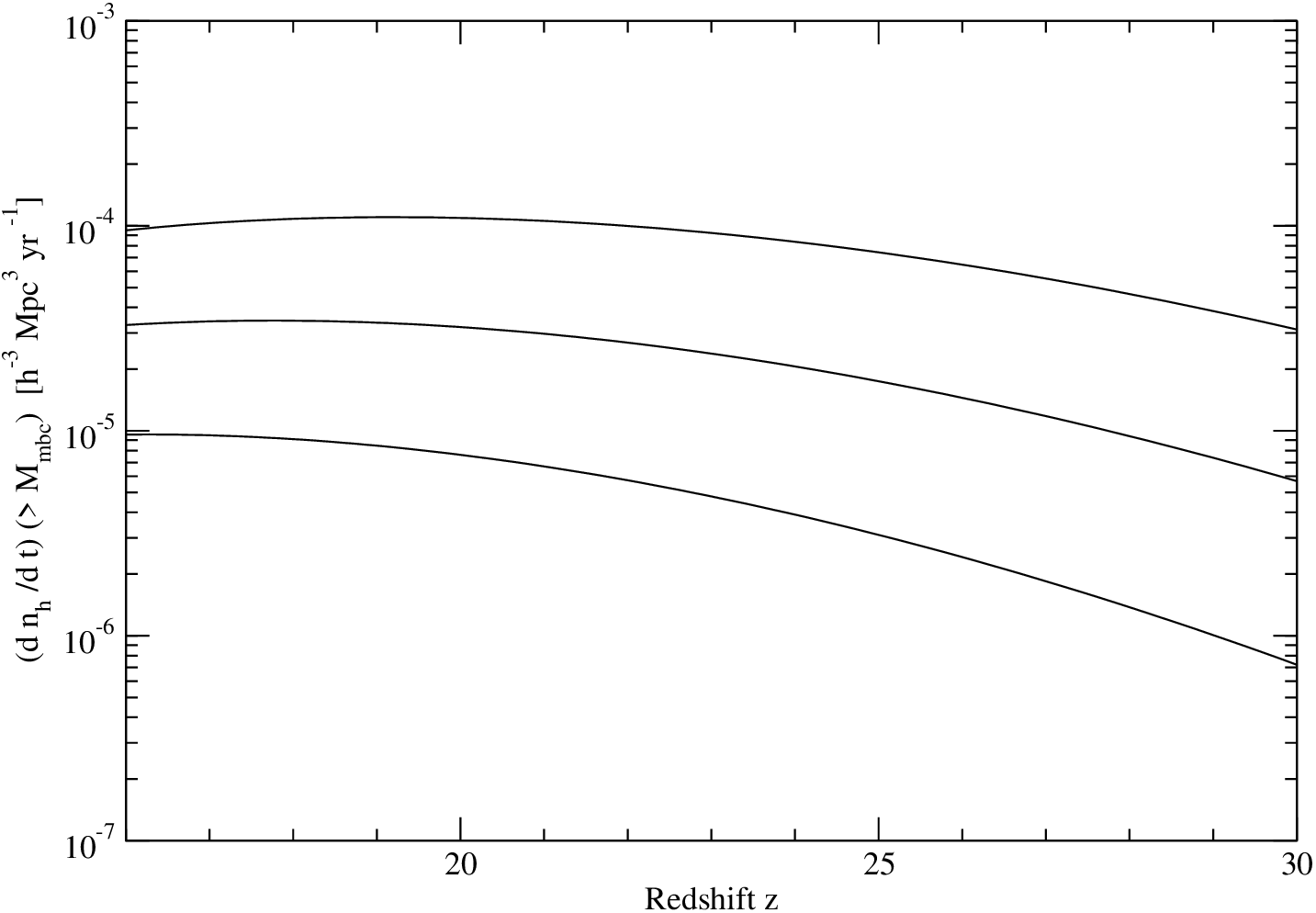}
  }
    \caption{{\it Top}: Redshift dependent comoving number density of haloes
      with mass $M_h > M_{min}$. From top to bottom the curves
      correspond to a $M_{min} = \{1,3,10\}\times 10^5 M_{\odot}
      h^{-1}$. {\it Bottom}: Rate of change of minihalo
       number density. The curves are in the same order.
    }
    \label{fig:Mmbc_vs_z}
\end{figure*}

If we identify a GRB event with the collapse of a minihalo, the rate of
GRBs at any given redshift is then approximately
\begin{equation}
  \frac{\d \nu_{GRB}}{\d z} \d z \approx 4 \pi D(z)^2 \d D(z)
    \int_{M_{min}}^{\infty}\frac{\d^2
    n_h}{\d t ~\d M} \d M
\end{equation}
where we have used the rate of change of the minihalo mass function
\begin{eqnarray}
  \label{eq:grbrate}
  \frac{\d^2 n_h}{\d t ~\d M_h} &=& \left(\frac{2}{\pi}\right)^{1/2} \frac{\rho_0}{M_h^2}
  \left|\frac{\d \ln \sigma}{\d \ln M_h}\right| ~u \nonumber \\
  & & \times ~\Exp\left[{-\frac{1}{2}
  u^2}\right]\left[\frac{1}{u} \frac{\d u}{\d t} - u \frac{\d u}{\d t}\right] \\ \nonumber
  &=& \frac{\d n_h}{\d M_h} \frac{\d u}{\d t} \left[\frac{1}{u} - u \right]
\end{eqnarray}
and $u = \delta_c/[\sigma(M_h) * D_{grow}(t)]$.
We have shown the mass function and its rate of change in figure \ref{fig:Mmbc_vs_z}.

\subsection{Rate of GRBs received}
For GRBs at the very high redshifts we are considering, no afterglow
emission can be detected anymore. An accurate redshift determination
is thus not possible unless a unique distance-luminosity relation
for GRBs could really be established.
If all GRBs do have the same total luminosity and peak photon flux,
we have a unique distance-luminosity relation and thus a unique
redshift-luminosity relation. Eqn.~(\ref{eq:grbrate})
above can then be converted into a flux distribution $F_{GRB}(z)$
using $\d z \propto -2\pi D(z)^3/(10^{52 - 54} ~{\rm erg}) (\d D/\d
z)^{-1} \d F_{GRB}$.

However, the case for a unique GRB distance-luminosity relation has
yet to be made. To test our model we therefore focus on a prediction
of the rate of GRBs to be compared with those detected by e.g.~the {\it
  SWIFT} mission.
We integrate eq. (\ref{eq:grbrate}) over the redshift range of
population III star formation to arrive at the total rate of GRBs
received.

\begin{equation}
  \nu_{GRB,rec} = \int_{z_{max}}^{z_{min}} (1 + z)^{-1} \frac{\d
  \nu_{GRB}}{\d z} \d z
\end{equation}

\begin{figure}
   \includegraphics[width=8.5cm]{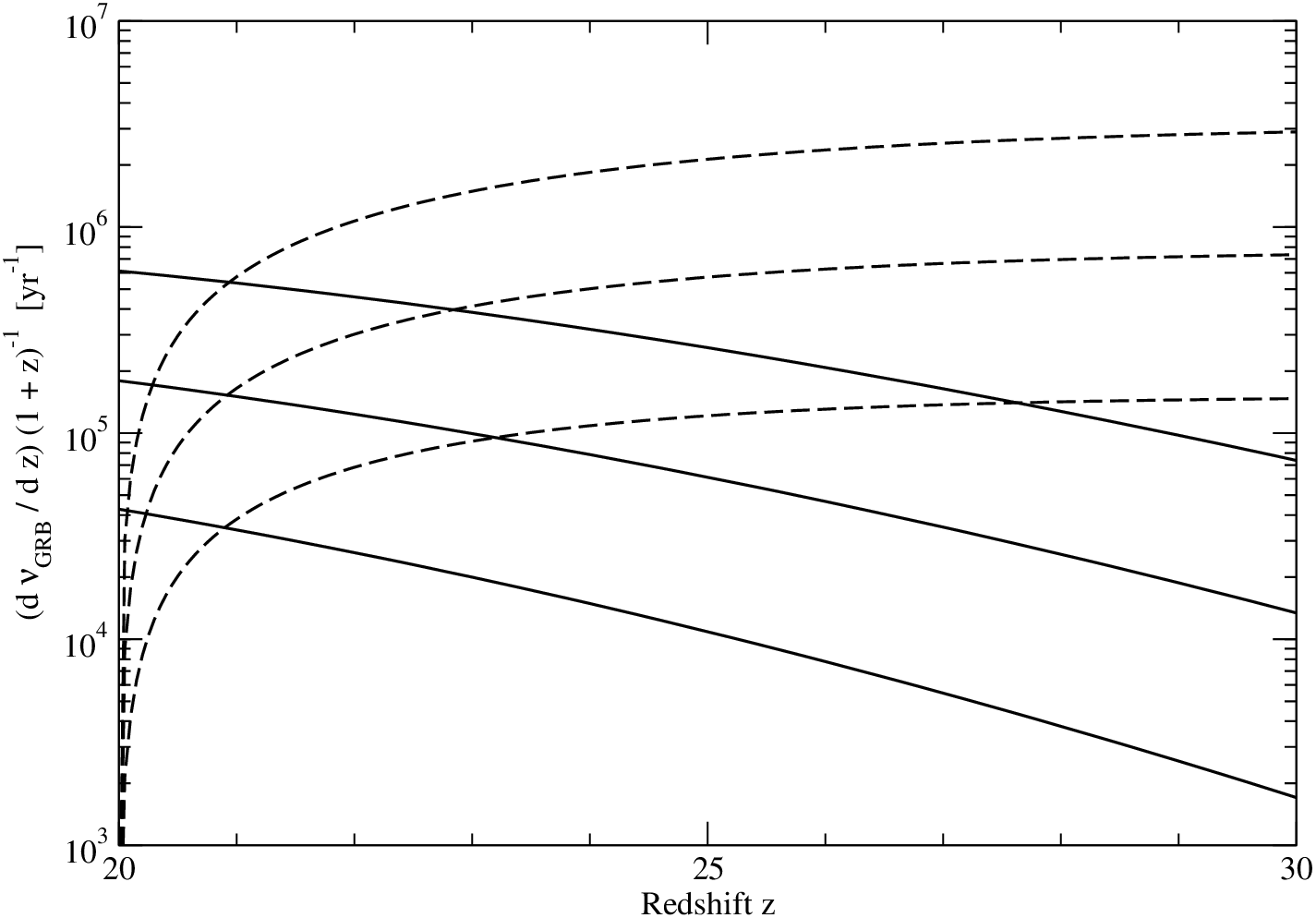}
    \caption{
      Number of GRB events received per year originating from redshift
      $z$ to $z + \d z$. The dashed curve shows the total number of
      GRB events from $z \geq 20$. From top to bottom the curves
      correspond to a $M_{min} = \{1,3,10\}\times 10^5 M_{\odot}
      h^{-1}$.
    }
    \label{fig:vGRB_z}
\end{figure}

This is shown in figure \ref{fig:vGRB_z}, together with rate of
GRB events originating from any given redshift interval for three
different values of $M_{min}$, assumed constant over the redshift
range $20 < z < 30$. We find that the total number of GRB events
received per year from this redshift range varies between
$\nu_{GRB} \approx 3 \times 10^6 - 2\times 10^5 {\rm yr}^{-1}$ for
$M_{min} = 10^{5}$ to $10^{6}$ \Msun\ respectively. Even for
$M_{min} = 3\times 10^5$ \Msun\ at $z = 25$, it may still make
more sense to consider a larger halo mass, since by definition
baryons -- and therefore stars -- in minihaloes only cool on a
Hubble timescale, whereas we have assumed that GRB events coincide
with the collapse of the haloes within which they occur. At $z =
25$, for instance, a halo mass three times larger than the
minihalo mass we used would result in a cooling time that is about
a tenth of the Hubble time at that redshift and so more in line
with this assumption.

A conservative estimate for the total number of GRB events received is
$\nu_{GRB} \approx 2 \times 10^5 {\rm yr}^{-1}$, assuming that halo
collapse, star formation and GRB coincide and that first star
formation ends at $z = 20$. How many of these could actually be
observed depends on whether or not GRBs are luminous enough to still
be detected at these distances.

\subsection{Gamma rays from MBH mergers}
If a small fraction of the gravitational energy from MBH mergers
is released in the form of gamma rays - maybe due to the fast
accretion of large amounts of material in the very close vicinity
of the MBHs - we may be able to see these. For an inspiralling MBH
binary the energy radiated away in gravitational waves, $L_{gw}$
is \cite{shapiro83}
\begin{eqnarray}
  L_{gw} &=& \frac{32}{5}\frac{G^4}{c^5}\frac{M_{12}^3 ~\mu_{12}^2}{a_0^5}
  \nonumber \\
  &=& 5.43
  \times10^{40} \left(\frac{a_0}{10^8 {~\rm
  m}}\right)^{-5}\left(\frac{M_{12}}{10^3 ~{\rm M}_{\odot}}\right)^3
  \nonumber \\
  & & \times \left(\frac{\mu_{12}}{10^3 ~{\rm M}_{\odot}}\right)^2 ~{\rm erg ~s}^{-1}
\end{eqnarray}
and combining this with the characteristic merger time scale eq
\ref{eq:GWinspiraltime} this becomes
\begin{eqnarray}
  L_{gw} &=&2.13\times10^{50} \left(\frac{\tau_0}{{\rm s}}\right)^{-5/4}
  \left(\frac{M_{12}}{10^3 ~{\rm M}_{\odot}}\right)^{1/2} \nonumber \\
  & & \times\left(\frac{\mu_{12}}{10^3 ~{\rm M}_{\odot}}\right)^{3/4} ~{\rm erg ~s}^{-1}
\end{eqnarray}
If we now multiply this with $1/f_{max}$ (eq \ref{eq:GWfmax}) we
can estimate the energy radiated away on the last orbit before
final coalescence.
\begin{equation}
   L_{gw} \approx 3\times10^{50}
  \left(\frac{M_{12}}{10^3 ~{\rm M}_{\odot}}\right)^{1/4}
  \left(\frac{\mu_{12}}{10^3 ~{\rm M}_{\odot}}\right)^{3/4} ~{\rm erg}
\end{equation}
It is interesting to compare this with GRBs. A typical GRB lasts
between a fraction of a second to minutes and emits total energies
between $10^{51} - 10^{54}$ erg \cite{piran00} and about a factor
100 less if we assume beamed emission. As far as the duration of
GRBs is concerned this is matched by the period of the last orbits
of MBH binaries with mass $M_{12}\sim 1000$ \Msun ($\tau_0 =
1/f_{max} \sim 0.1$ seconds) to $M_{12}\sim 10^6$ \Msun ($\tau_0
\sim 100$ seconds). This means that if some fraction
$\epsilon_{\gamma} \grtsim 3\times10^{-2}$ of gravitational wave
energy is emitted in gamma rays, merging MBHs could account for
some beamed GRBs. At the lower end the detection limit of the {\it
BATSE} instrument on the {\it Compton Gamma Ray Observatory}
\footnote{{\it BATSE}'s burst sensitivity is quoted as
$3\times10^{-8}
  {\rm erg~cm}^{-2}$.}
implies $\epsilon_{\gamma} \grtsim 10^{-6}$ for MBH mergers to be
detected out to cosmological distances. For isotropically emitting
GRBs $\epsilon_{\gamma}$ would need to be a factor 100 or so
higher.

\section{Summary and Conclusions}
\label{sec:summary}
In this paper we were mainly concerned with probing our model at high
redshifts. The merging of MBHs in the context of the hierarchical
merging of galaxies and haloes would lead to the emission of
gravitational waves, that could be detected with the LISA mission due
to be launched in the next decade.
We can draw the following conclusions:
\begin{itemize}
\item The merger rate per unit redshift peaks at $z \sim 4$--$6$. The
  peak shifts to lower redshifts if events with higher dimensionless strain
  amplitudes are considered.
\item The predicted merger events all have strain amplitudes above
  $h_c > 10^{-23}$, and most above $10^{-22}$, so that LISA should be
  able to detect most of these.
\item The total number of MBH merger events received are about $10^4
 $--$10^5$ per year. This is at least an order of magnitude larger than
  typical estimates for (`major') mergers of SMBHs in hierarchical structure
  formation scenarios. Our larger number arises from the
  inclusion of a lot of mergers with fairly large binary mass ratios,
  i.e. what we may qualify as `minor' mergers. The required
  population of MBHs are the (merged) remnants of the first
  massive stars in the universe. An estimate of this population
  was the subject of paper I in this series.
\item In contrast to previous analytical work our predictions are
  based on a more accurate and explicit modelling of the
  dynamical evolution of MBHs in galactic haloes. This has allowed us
  to follow MBH merging in haloes at higher redshift and therefore also
  down to lower precursor masses.
\item If the collapse of the first massive population III stars into
  MBHs is accompanied by gamma-ray bursts (GRBs), these might allow us to probe
  the abundance of the first stars as well as their spatial
  distribution. We predict $10^5$--$10^6$ GRBs per year originating
at redshifts higher
  than 20, if there is one massive star per star-forming minihalo
  associated with a GRB.
\end{itemize}

\begin{itemize}
\item A fraction of the gravitational wave
  energy in MBH mergers may be released in gamma rays, e.g. through
  interaction with remnant gas in the close vicinity of the
  coalescing binary. If GRB emissions are beamed this fraction must be $\epsilon_{\gamma} > 10^{-6}$
  for the resulting bursts of gamma
  rays to be detectable at cosmological distances and $\epsilon_{\gamma} > 3\times10^{-2}$ to possibly account for some of the proper
  GRBs. However, the short event duration makes it difficult to
  identify them unless a GRB style after-glow is associated with them.
\end{itemize}

While the prediction of gravitational wave events and GRBs both
depend sensitively on the initial assumption of massive star
formation, the additional assumptions of efficient merging for
gravitational waves and that GRBs are created in the process of
massive star collapse, further increase uncertainty. All these
assumptions are in principle verifiable primarily with improved
numerical simulations (The role of gas in BH binary
  mergers, improved resolution in star formation and collapse simulations in a
  cosmological context). For the time being, the prediction of event
rates of 10 to 100 a day should provide motivation to look out for
them in present and upcoming observations. Qualitatively the most
important thing is to note, that we could potentially observe the
highly energetic events accompanying both the formation (via GRBs)
and merger (by gravity waves) of MBHs, and therefore probe the
high redshift regime.

\section*{Acknowledgements}
The authors wish to thank R.~Bandyopadhyay, G.~Bryan, J.
Magorrian and H.-W.~Rix for helpful discussions.
RRI acknowledges support from Oxford University and St.~Cross College, Oxford.
JET acknowledges support from the Leverhulme Trust and from the Particle
Physics and Astronomy Research Council (PPARC).


\end{document}